\documentclass[a4paper,twocolumn,english,aps,longbibliography,reprint]{revtex4-1}
\usepackage[T1]{fontenc}
\usepackage[latin9]{inputenc}
\usepackage{color}
\usepackage{babel}
\usepackage{bm}
\usepackage{amsmath}
\usepackage{graphicx}
\usepackage[unicode=true,pdfusetitle,
 bookmarks=false,
 breaklinks=false,pdfborder={0 0 1},backref=false,colorlinks=true]
 {hyperref}
\hypersetup{
 citecolor=blue,urlcolor=blue}
\usepackage{breakurl}

\makeatletter

\usepackage{babel}
\usepackage{babel}
\usepackage{babel}
\usepackage{babel}
\usepackage{babel}
\usepackage{color}
\usepackage{bm}

\usepackage{breakurl}

\addto\captionsenglish{}

\makeatother

\begin{document}

\preprint{APS/123-QED}

\title{Zero-line modes at stacking faulted domain walls in multilayer graphene}

\author{Changhee Lee$^{1}$}

\author{Gunn Kim$^{2}$}

\email{gunnkim@sejong.ac.kr}

\author{Jeil Jung$^{3}$}

\email{jeil.jung@gmail.com}

\author{Hongki Min$^{1}$}

\email{hmin@snu.ac.kr}

\affiliation{$^{1}$ Department of Physics and Astronomy, Seoul National University,
Seoul 08826, Korea}

\affiliation{$^{2}$ Department of Physics and Graphene Research Institute, Sejong
University, Seoul 143-747, Korea}

\affiliation{$^{3}$ Department of Physics, University of Seoul, Seoul 02504,
Korea}
\begin{abstract}
Rhombohedral multilayer graphene is a physical realization of the
chiral two-dimensional electron gas that can host zero-line modes
(ZLMs), also known as kink states, when the local gap opened by inversion
symmetry breaking potential changes sign in real space. Here we study
how the variations in the local stacking coordination of multilayer
graphene affects the formation of the ZLMs. Our analysis indicates
that the valley Hall effect develops whenever an interlayer potential
difference is able to open up a band gap in stacking faulted multilayer
graphene, and that ZLMs can appear at the domain walls separating
two distinct regions with imperfect rhombohedral stacking configurations.
Based on a tight-binding formulation with distant hopping terms between
carbon atoms, we first show that topologically distinct domains characterized
by the valley Chern number are separated by a metallic region connecting
AA and AA$'$ stacking line in the layer translation vector space.
We find that gapless states appear at the interface between the two
stacking faulted domains with different layer translation or with
opposite perpendicular electric field if their valley Chern numbers
are different. 
\end{abstract}
\maketitle

\section{Introduction}

Rhombohedral graphene multilayers may host one-dimensional metallic
states at the domain wall between two broken inversion symmetry insulating
regions with opposite mass signs either due to reversal of sign of
the perpendicular electric field \citep{Martin2008,Jung2011,Bi2015,Zhang2013,DeSena2014}
or reversal of stacking order \citep{Semenoff2008,Vaezi2013,Zhang2013,Jung2012}.
As a rule of thumb, at the interface between two domains made by $N$-layer
rhombohedral graphene, the co-propagating $N$ gapless metallic states
appear in each valley inside the bulk gap along the interface \cite{Jung2011,Volovik2003,Yao2009},
whose origin can be traced back to the valley Hall effect \citep{Bi2015,Jung2011,Jung2012,Vaezi2013,Zhang2013}
associated with opposite Hall conductivities at each valley \citep{Xiao2007}.
Theoretically it was predicted that the zero-line modes (ZLMs), also
known as \emph{kink states} \cite{Jung2011,KILLI2012,Vaezi2013},
will have exceptional transport properties such as suppressed backscattering,
zero bend resistance, and chirality encoded current filtering \citep{Qiao2011,Qiao2014},
and have been proposed as splitters of Cooper pairs injected by superconducting
electrodes \citep{Schroer2015}. There has been recently a number
of experimental advancements related to the ZLMs in bilayer graphene.
The enhanced electron density signals at tilted boundaries due to
stacking faults in bilayer graphene were visualized through transmission
electron microscopy measurements \citep{Alden2013}, confirming the
theoretical expectations of finding stacking-dependent ZLMs at the
stacking domain walls \citep{Vaezi2013}. Recently a direct observation
of the ZLMs at AB/BA tilted boundaries in bilayer graphene through
infrared nanoscopy and subsequent measurement of ballistic transport
has been reported \citep{Ju2015}. Similar ZLMs have also been tailored
in devices with electric field domain walls by applying bias potentials
of opposite sign at different domains laying the first stepping stone
towards the systematic realization of ZLM devices based on bilayer
graphene \citep{Li2015}. While experimental studies of ZLM physics
are still in their infancy, new advancements in the study of ZLMs
in graphene are expected to follow in the near future.

In this article we investigate theoretically how the local deviations
of the stacking order from perfect rhombohedral stacking can impact
the band gap and valley Hall conductivities that underlie the formation
of the ZLMs in AB bilayer and ABC trilayer graphene based devices.
This is a practically relevant question in view of the relatively
small layer sliding energy barrier on the order of 4~meV in bilayer
graphene systems existing between the two lowest energy stacking AB
and BA configurations \citep{Jung2014moire,Alden2013}, making it
feasible to form sizable regions of misaligned stacking order due
to the presence of local defects, strain fields, and twist angles.
While the conditions of band gap opening for commensurate bilayer
geometries of arbitrary stacking had been discussed in Ref. \citealp{Park2015},
their influence in the formation of the ZLMs remains unexplored. Here
we show that the opening of a gap in stacking faulted multilayer graphene
is a sufficient condition for developing a finite valley Hall conductivity
that is presupposed in the formation of the ZLMs. This finding enables
us to identify the local stacking configuration phase space that defines
the topologically distinct domains necessary for the emergence of
the metallic gapless states at their interface walls, confirming the
intuitive expectation that the metallic states can also emerge at
the interface between the two domains with layer-translated stacking
structures if they are insulating and topologically distinct.

The article is structured as follows. We begin in Sec.~II by presenting
the model and theoretical background, discussing the tight-binding
Hamiltonian for layer-translated stacking configurations. We then
discuss in Sec.~III the conditions for the appearance of the ZLMs
at the domain walls in the presence of stacking faults. In Sec.~IV
we provide a more detailed analysis of the gap size, the critical
electric fields required for opening a band gap, and the influence
of the domain wall sharpness and field strength on the spreading width
of ZLMs.

\section{Model}

\begin{figure}[h]
\includegraphics[width=8cm]{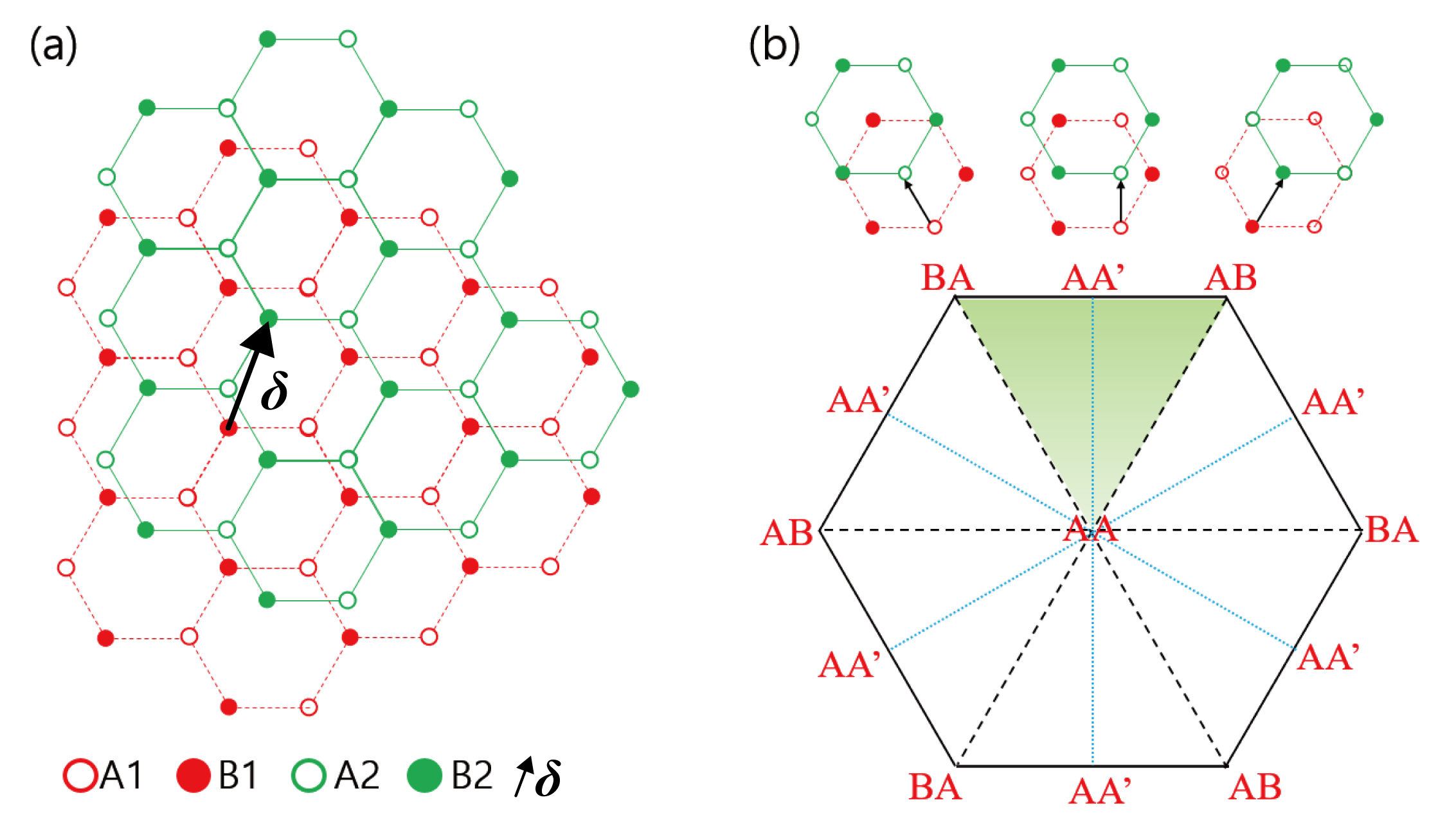} \caption{(a) Bottom (red) and top (green) layers and the translation vector
$\bm{\delta}$ represented with a black arrow. (b) Representation
of the different commensurate stacking configurations that can be
reached by in-plane translation. The three sets of red and green hexagon
pairs represent the BA, AA$'$, and AB stacking of bilayer graphene
achieved by sliding from AA stacked configuration into different directions. }
\label{fig:1} 
\end{figure}

In order to consider the effects of finite translation in multilayer
graphene band structure we begin by the simplest example of bilayer
graphene where the position of the top layer with respect to the bottom
layer is represented by a translation vector $\bm{\delta}$, as shown
in Fig.~\ref{fig:1}(a). If we set $\bm{\delta}=(0,0)$ for the AA
stacking structure, then $\bm{\delta}=(\pm\frac{\sqrt{3}}{6},\frac{1}{2})a$
correspond to the AB and BA stacking structures, respectively, while
$\bm{\delta}=(0,\frac{1}{2})a$ corresponds to AA$'$ stacking structure
\citep{Park2015}, where $a=2.46$ $\textrm{\AA}$ is the lattice
constant. Due to the periodic structure of bilayer graphene, the layer
translation vector space can be divided into triangular irreducible
zones, as shown in Fig.~\ref{fig:1}(b) with several representative
stacking structures. Similarly, we can easily generalize from the
bilayer graphene to multilayer cases. For multilayer graphene, we
assume for simplicity that each layer is translated along the same
direction and the relative translations between the adjacent layers
are the same.

The electronic band structure of multilayer graphene nanoribbon (MGNR)
geometries with one-dimensional band dispersions offers a convenient
platform for representing the dispersion of the ZLMs. To describe
band structures of the layer translated MGNRs, we use a tight-binding
method taking into account remote hopping terms up to the 15 nearest-neighbor
unit cells to fully account for contributions from all arbitrarily
displaced atoms and at the same time to evolve the Hamiltonian smoothly
with the layer translation.

The Hamiltonian in the second-quantized form reads 
\begin{eqnarray}
H & = & \sum_{l=1}^{N}H_{l}^{(0)}+\sum_{\langle l,l'\rangle}V_{ll'},\label{eq:tight_binding}\\
H_{l}^{(0)} & = & \sum_{i}U_{l,i}c_{l,i}^{\dagger}c_{l,i}+\sum_{\langle i,j\rangle}t_{l}(\bm{r_{i}}-\bm{r_{j}})c_{l,i}^{\dagger}c_{l,j},\nonumber \\
V_{ll'} & = & \sum_{i,j}t_{ll'}(\bm{r_{i}}-\bm{r_{j}})c_{l,i}^{\dagger}c_{l',j}\nonumber 
\end{eqnarray}
where $c_{l,i}^{\dagger}$ ($c_{l,i}$) corresponds to the creation
(annihilation) operator for an electron on the $i$-th site in the
$l$-th layer. The first term $H_{l}^{(0)}$ represents the single-layer
graphene Hamiltonian for the $l$-th layer with on-site potential
energy $U_{l,i}$ and intralayer hopping $t_{l}$ for each $l$-th
layer. The second term $V_{ll'}$ describes the interlayer coupling
between the $l$-th layer and $l'$-th layer with interlayer tunneling
$t_{ll'}$. Note that the index $i$ represents only the site of each
atom since spin does not play any role in our demonstration. As an
approximation we consider the interlayer hopping terms only between
adjacent layers. To fully account for contribution between arbitrarily
displaced atoms, we include remote hopping terms beyond the nearest-neighbor
approximation with the exponentially decaying form with distance,
\begin{equation}
t_{ll'}\left(\bm{d}\right)=V_{pp\pi}^{0}e^{-\frac{d-a_{0}}{r_{0}}}\sin^{2}\theta+V_{pp\sigma}^{0}e^{-\frac{d-d_{0}}{r_{0}}}\cos^{2}\theta\label{eq:remote_hopping}
\end{equation}
where $V_{pp\pi}^{0}\approx-2.7$ eV and $V_{pp\sigma}^{0}\approx0.48$
eV represent the nearest-neighbor intralayer and interlayer hopping
terms at the distance $a_{0}$ and $d_{0}$, respectively, where $a_{0}=a/\sqrt{3}\approx1.42$
$\textrm{\AA}$ is the in-plane carbon-carbon distance and $d_{0}=3.35$
$\textrm{\AA}$ is the interlayer separation. Here $\bm{d}$ is a
displacement vector between two carbon atoms and $\theta$ is the
angle between the $z$ axis and $\bm{d}$. Following Ref.~\citealp{Koshino2013},
we take $r_{0}=0.453$ $\textrm{\AA}$.

The valley Chern number associated with each gapped Dirac cone, defined
as the Chern number for a single valley, allows us to count the number
of expected ZLMs in the presence of a domain wall. For example, for
rhombohedral stacked $N$-layer graphene, the valley Chern number
can be easily estimated from the low energy effective theory given
by 
\begin{equation}
H=\left(\begin{array}{cc}
-\frac{U}{2} & -\gamma(\tau_{z}\pi_{x}-i\mu\pi_{y})^{N}\\
-\gamma(\tau_{z}\pi_{x}+i\mu\pi_{y})^{N} & \frac{U}{2}
\end{array}\right),\label{eq:chiral_gas}
\end{equation}
where $(\pi_{x},\pi_{y})=\frac{\hbar v_{{\rm F}}}{\gamma}(k_{x},k_{y})$
with $\frac{\hbar v_{{\rm F}}}{a}=-\frac{\sqrt{3}}{2}V_{pp\pi}^{0}$
and $\gamma=V_{pp\sigma}^{0}$\cite{Min2008a,Min2008b}. The corresponding
valley Chern number is given by $\mathcal{C}=\frac{N}{2}\tau_{z}\mu\,{\rm sgn}(U)$,
where $\tau_{z}=\pm1$ is the valley index and $\mu=\pm1$ represents
the direction of rhombohedral stacking. {[}For example, in bilayer
graphene $\mu=\pm1$ corresponds to AB (BA) stacking.{]} Note that
the valley Chern number is proportional to the valley index $\tau_{z}$
and the sign of $U$, and its sign is flipped if we reverse the stacking
sequence. Thus the valley Chern numbers change $\pm N$ across the
interface between two domains with opposite $\mu$ or $U$. Although
the validity of this effective model is limited to low energies near
the Fermi energy, it is useful for illustrating the topological nature
of the valley Chern number in rhombohedral $N$-layer graphene that
should lead to $N$ gapless metallic states in each valley in the
presence of domain walls with opposite stacking order or perpendicular
electric field direction.

\begin{figure}[h]
\includegraphics[width=1\linewidth]{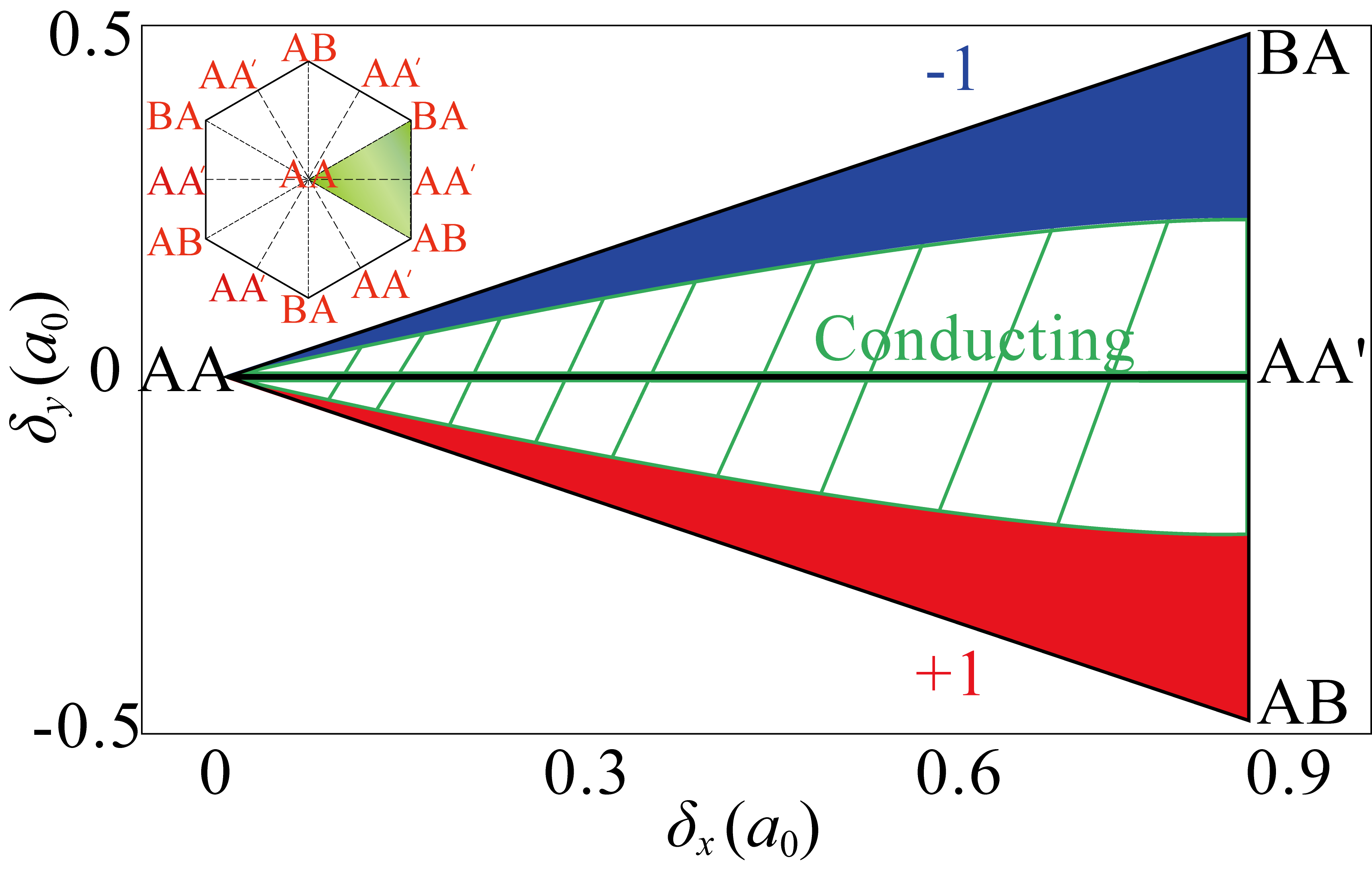} \caption{Chern number phase space map at the \emph{K} valley of bilayer graphene
under the finite potential difference $U=0.5$ eV between layers.
The valley Chern numbers are obtained by integrating the Berry curvature
around a valley of the gapped four-band low energy effective Hamiltonian
of stacking faulted bilayer graphene in Eq.~(\ref{eq:A1}) and have
opposite sign in the opposite valley. Note that because of the topological
nature of the valley Chern numbers, the same values can be obtained
if we use the lattice model in Eq.~(\ref{eq:tight_binding}). The
region with oblique lines represents stackings in metallic phase under
the applied potential difference in which the valley Chern number
cannot be defined. The hexagonal stacking map is rotated by $30\protect\textdegree$
for representation convenience.}
\label{fig:2} 
\end{figure}

When we introduce an in-plane translation in a gapped phase we expect
that the valley Chern number, which characterizes the topological
property of the band structure in the continuum limit around a valley,
will remain unchanged until the energy bands near the Fermi level
touch each other. Our explicit calculation of the Chern number through
integration of the Berry curvature around each valley confirms that
indeed this is the case, as we illustrate in the Chern number phase
space map in Fig.~\ref{fig:2} for which we use a constant interlayer
potential difference $U=0.5$ eV. The distinct insulating regions
represented in Fig.~\ref{fig:2} in red (blue) with the valley Chern
numbers $+1$ ($-1$) at the \emph{K} valley have dominantly AB (BA)
stacking and are divided by the line connecting the AA and AA$'$
stacking configurations. The system remains invariably metallic at
this division line even if the phase map changes depending on the
magnitude of $U$. We will show that the ZLMs appear even in stacking
faulted systems whenever there are two insulating domains with opposite
Chern numbers, resulting either from opposing stacking regions (AB/BA-like)
or electric field directions.

\section{zero-line Modes between stacking faulted domains}

\begin{figure}[h]
\includegraphics[width=1\linewidth]{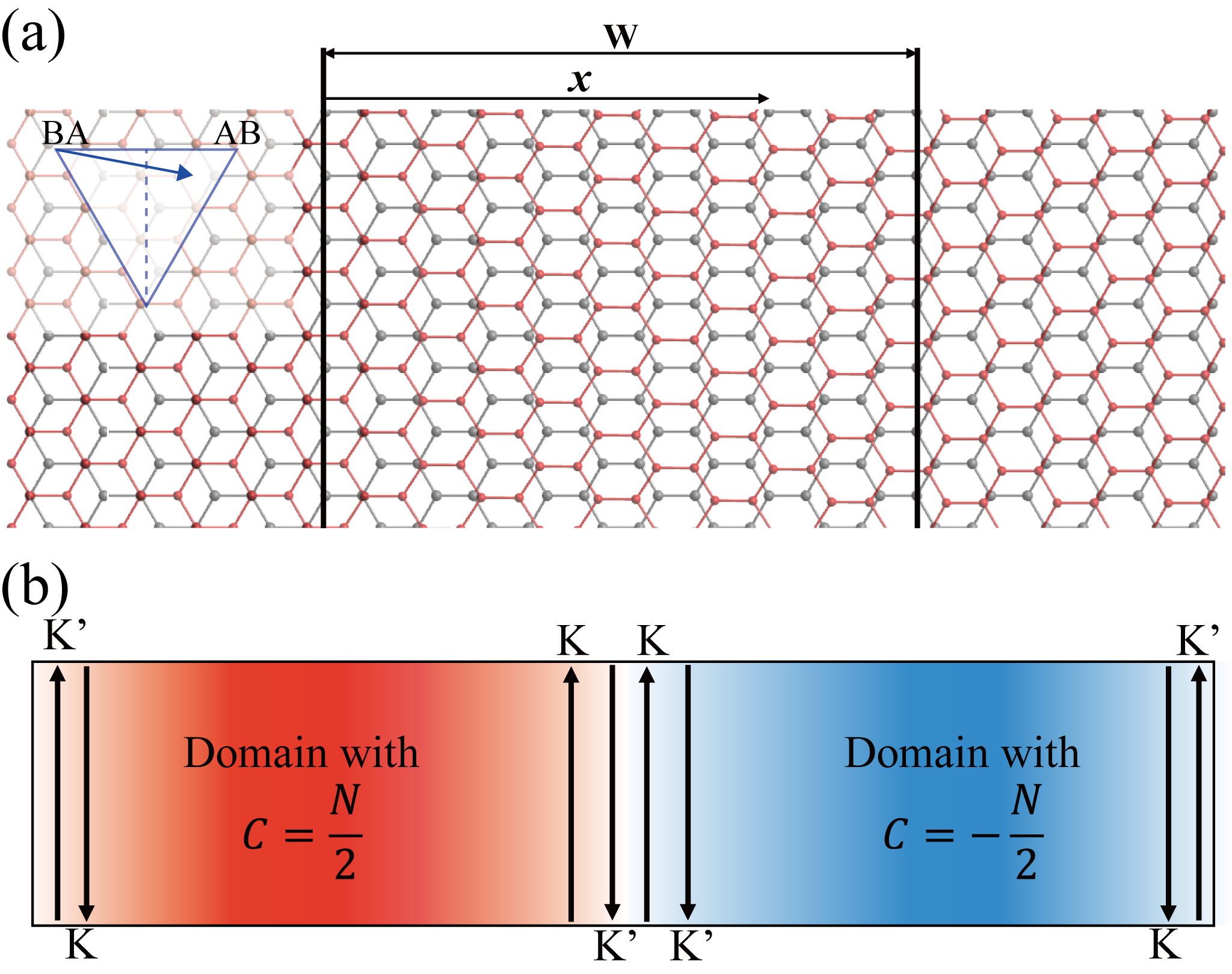} \caption{(a) Illustration for a zigzag terminated 2GNR-LSDW with the domain
wall size $W$. The carbon atoms in the top layer are depicted in
red and those in the bottom layer are depicted in black. The inset
indicates an irreducible zone. The blue arrow in the inset is the
trace of the head of displacement vector field from the left domain
to the right domain. Note that AA$'$-like stacking appears transiently
in the middle of domain wall. (b) Schematic illustration of valley-polarized
ZLMs between zigzag terminated domains with the valley Chern number
$\mathcal{C}=\frac{N}{2}$ in one domain and $\mathcal{C}=-\frac{N}{2}$
in the other domain at the valley \emph{K}. For armchair terminated
nanoribbons, edge states localized on outer edges do not appear. }
\label{fig:3} 
\end{figure}

We study the impact of the stacking faults in the ZLMs by analyzing
their real-space probability distribution and the energy band structure.
The first case to consider is the layer stacking domain wall (LSDW)
\citep{Zhang2013} in a bilayer graphene nanoribbon (2GNR) under the
influence of a constant interlayer potential difference. Figure~\ref{fig:3}
schematically illustrates a 2GNR with a LSDW constructed by sliding
the carbon atoms continuously as 
\begin{equation}
\bm{\delta}(x)=\begin{cases}
\bm{\delta}_{1} & (x<0,\mbox{ domain 1})\\
\bm{\delta}_{1}+\frac{\bm{\delta}_{2}-\bm{\delta}_{1}}{W}x & (0\le x\le W,\mbox{ domain wall})\\
\bm{\delta}_{2} & (x>W,\mbox{ domain 2})
\end{cases}\label{eq:4}
\end{equation}
where $\bm{\delta}_{1}$ and $\bm{\delta}_{2}$ are layer translation
vectors in domains 1 and 2, respectively and $x$ is distance from
the starting point of the domain wall. The domain wall size $W$ is
chosen to be about 10 nm to meet the observed domain wall length \cite{Butz2014,Ju2015,Lin2013},
and the size of each domain is set as 136 nm in our calculation. A
different choice of displacement vector field $\bm{\delta}(x)$, for
example, a $\tanh x$ variation along the displacement direction,
does not change the result qualitatively. The effect of the domain
wall size will be discussed later.

\begin{figure*}[htb]
\includegraphics[width=1\linewidth]{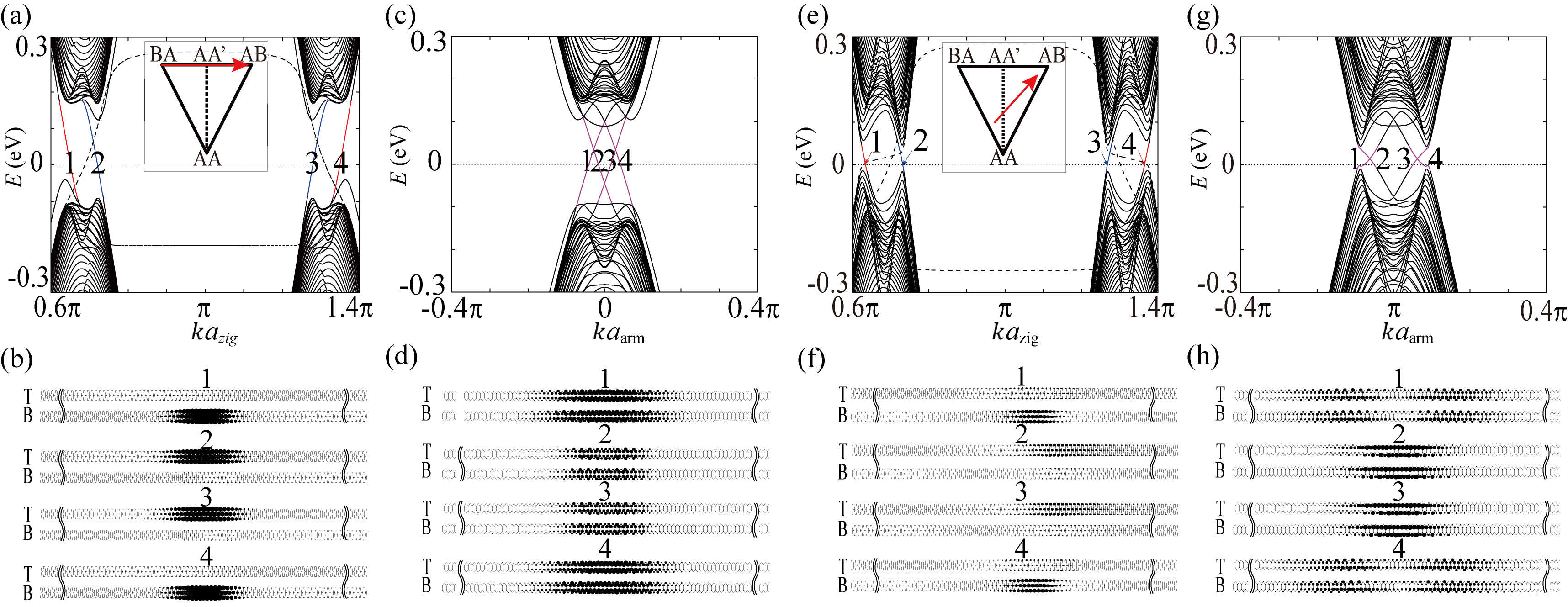} \caption{Band structures of various 2GNR-LSDW and corresponding probability
distributions for ZLMs. (a) and (b) The band structure and probability
distribution for the zigzag terminated 2GNR-LSDW with AB-BA, respectively
whereas (e) and (f) are for the zigzag terminated 2GNR-LSDW with two
arbitrarily stacked domains. (c) and (d) The band structure and probability
distribution for the armchair terminated 2GNR-LSDW with AB-BA, respectively,
whereas (g) and (h) are for the armchair terminated 2GNR\_LSDW with
two arbitrarily stacked domains. Insets in (a) and (e) indicate domains
of 2GNR-LSDW and transient atomic configuration in the domain wall.
Colored solid lines in the band structures are ZLMs and dashed lines
are edge states localized on outer edges of the nanoribbon. The modes
3 and 4 are the time reversal counter parts of the modes 2 and 1,
respectively. Here, $a_{{\rm zig}}=a$ and $a_{{\rm arm}}=\sqrt{3}a$
are the lattice constants of zigzag and armchair nanoribbons, respectively,
T (B) stands for the top (bottom) layer, and we use the interlayer
potential difference $U=0.5$ eV.}
\label{fig:4} 
\end{figure*}

The 2GNR in Fig.~\ref{fig:3} has a finite width along the horizontal
axis and extends infinitely along the vertical direction, which allows
us to use periodic boundary condition with a well-defined crystal
momentum $k$. In the middle of the ribbon, there is a domain wall
of width $W$ where the stacking structure changes smoothly from BA
to AB stacking. The arrow in the inset is the trace of the head of
displacement vector field showing how stacking changes across the
domain wall from the left to right domains.

\begin{figure}[h]
\includegraphics[width=1\linewidth]{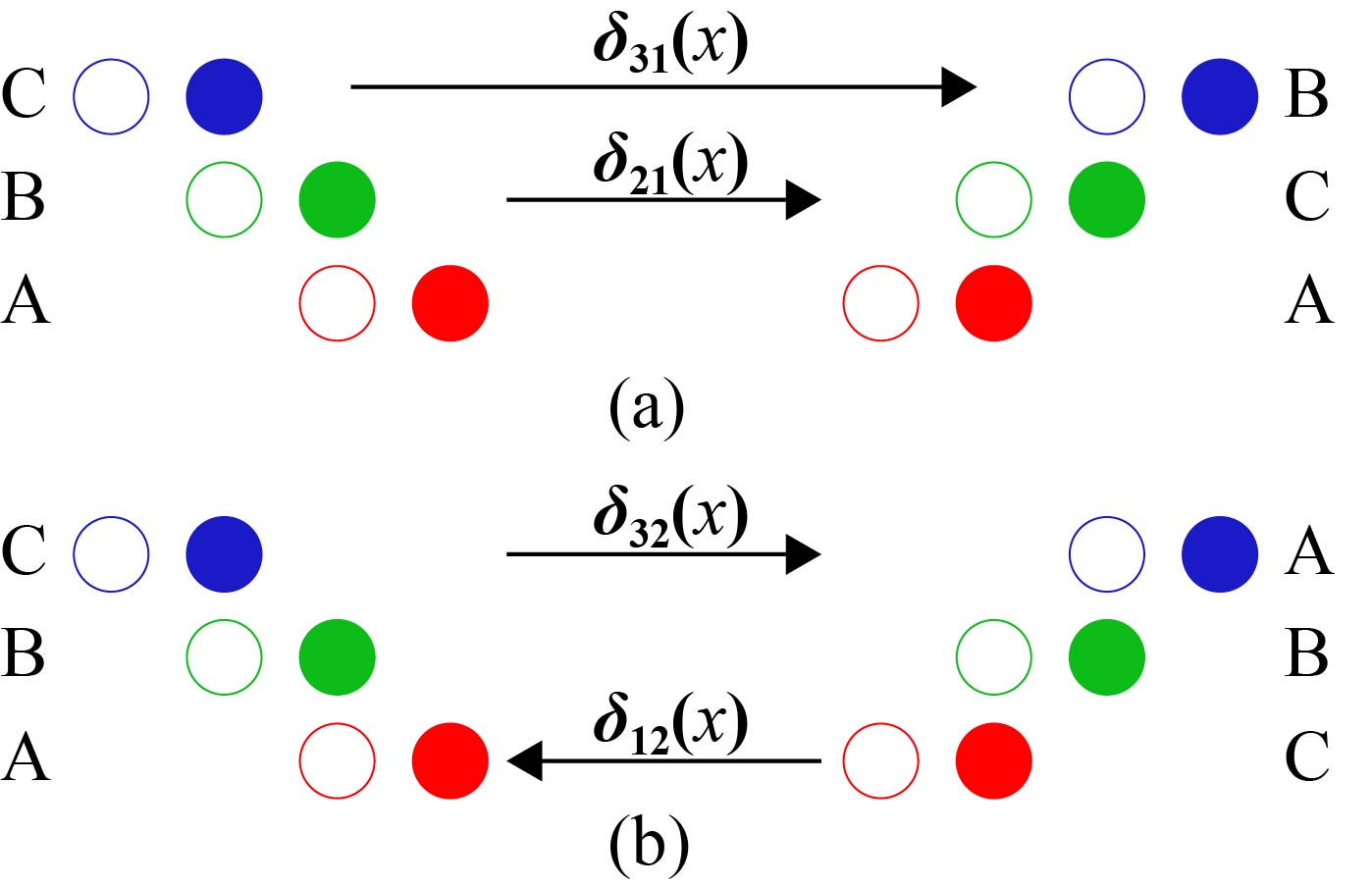} \caption{Schematic for layer translation in trilayer graphene. $\bm{\delta}_{ij}$
represents the displacement of atoms in the \emph{i}-th layer relative
to atoms in the \emph{j-}th layer. (a) In the ABC-ACB type domain
wall, the top layer is stretched twice than the middle layer. That
is, $\bm{\delta}_{31}=2\bm{\delta}_{21}$. (b) For ABC-CBA type, the
top layer is stretched while the bottom layer is contracted relative
to the middle layer. Here, $\bm{\delta}_{32}=-\bm{\delta}_{12}$.
The arrows represent the direction of the displacement of atoms in
each layer from their locations in the left domain. }
\label{fig:5} 
\end{figure}

\begin{figure*}[htb]
\includegraphics[width=1\linewidth]{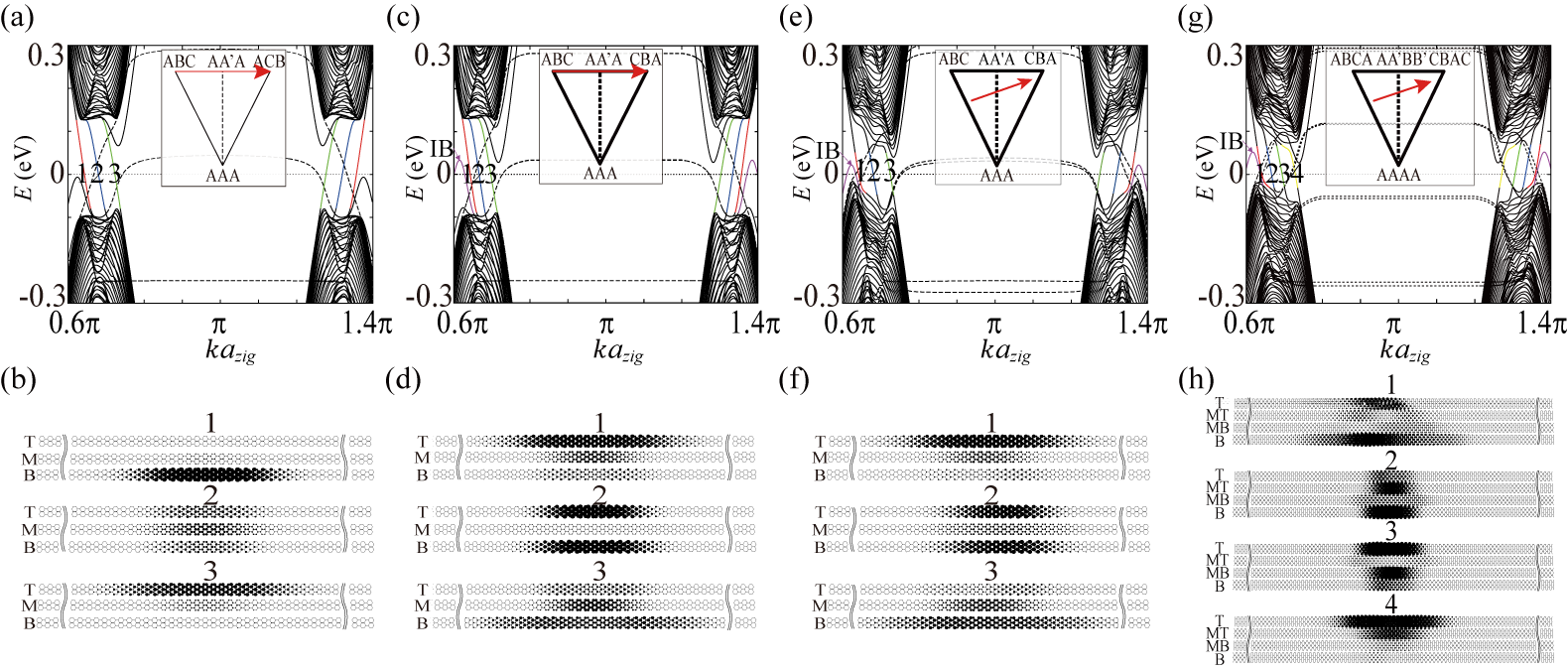} \caption{Band structures of various MGNR-LSDW and corresponding probability
distributions for ZLMs. (a) The band structure for ABC-ACB type 3GNR-LSDW,
(c) is for ABC-CBA type 3GNR-LSDW, (e) is for 3GNR-LSDW with two layer
translated domains with the atom displacement chosen to be ABC-CBA
type, and (g) is for the 4GNR-LSDW where the domain type is the extended
version of ABC-ACB type. (b), (d), (f), and (h) The probability distributions
of ZLMs in the corresponding band structures, respectively. Here we
only present modes in one valley and omit modes in other valley, which
are related with time-reversal symmetry. Interface bound modes crossing
the zero-energy are labeled as IB in (c), (e) and (g). M represents
the middle layer of trilayer graphene, whereas MT and MB represent
the middle-top layer and the middle-bottom layer in tetralayer graphene.
Here, $a_{{\rm zig}}=a$ is the lattice constants of zigzag and armchair
nanoribbons, respectively, and we use the interlayer potential difference
$U=0.5$ eV between the top and bottom layers. }
\label{fig:6} 
\end{figure*}

When we carry out explicit tight-binding calculations in the ribbon
geometries, we can observe that the ZLMs are indeed present even when
the stacking configurations at the domains have an in-plane shift.
In Fig.~\ref{fig:4} we show the band structures of 2GNRs with LSDW
and the corresponding probability distribution for states slightly
above the Fermi level for zigzag and armchair edge alignments. The
left two panels are for LSDW with AB and BA stacked domains at the
two sides whereas the right two panels are for LSDW with two domains
which have layer-translated stacking arrangements whose translation
vectors are separated by the AA-AA$'$ line, as seen in the insets
of Figs.~\ref{fig:4}(a) and \ref{fig:4}(e), respectively. First
consider the AB-BA nanoribbon with the zigzag arrangement in Figs.
\ref{fig:4}(a) and \ref{fig:4}(b). The dashed lines in Fig. \ref{fig:4}(a)
represent the edge state energy bands whose zero-energy states are
localized mainly on the outer edges \cite{Castro2008,Jung2011,Li2012,Delplace2011}.The
red and blue solid lines represent the energy bands whose zero-energy
states are localized around the domain wall, as seen by the size of
black filled circles in Fig. \ref{fig:4}(b). (Here we omitted the
edge states in the probability distributions because we focus on the
ZLMs.) Note that there are two dispersive ZLMs per valley, which can
be estimated by the valley Chern number difference between the two
sides, as shown in previous works \citep{Ju2015,Vaezi2013,Zhang2013,Jung2011}.
Importantly, these ZLMs still appear when the domains of 2GNR are
not in Bernal stacking as seen in Figs. \ref{fig:4}(e) and \ref{fig:4}(f).
Note that the probability distribution of the modes 3 and 4 are identical
to the modes 2 and 1, respectively, due to time-reversal symmetry.
A similar conclusion can be reached to armchair nanoribbons as seen
in Figs. \ref{fig:4}(c), \ref{fig:4}(d), \ref{fig:4}(g), and \ref{fig:4}(h),
in which edge states are absent. For the armchair nanoribbons, however,
the\emph{ K} and \emph{K'} valleys overlap in momentum space and the
ZLMs anticross leaving a small gap due to the mixing of the two valleys.
Our calculations support the fact that the presence of ZLMs are mainly
defined by the difference in the valley Chern numbers between the
left and right domains in the nanoribbon geometry studied.

For multilayers beyond bilayer graphene such as a trilayer, the analysis
becomes more complicated due to the diverse options for choosing a
fixed layer when creating a LSDW. For example, for trilayer graphene
nanoribbons (3GNRs), we can either fix the bottom layer while moving
the middle and top layers in the same direction or fix the middle
layer while moving the top and bottom layers in the opposite direction.
(As assumed in Sec. II, we will not consider the case with staggered
layer translation.) We distinguish these two options by naming ABC-ACB
and ABC-CBA, respectively, as seen in Fig.~\ref{fig:5}. Here $\bm{\delta}_{ij}$
represents the displacements of atoms in the $i$-th layer relative
to the atoms in the $j$-th layer. For the ABC-ACB type LSDW, atoms
in each layer are displaced in accordance with $\mathbf{\bm{\delta}}_{31}=2\,\bm{\delta}_{21}$,
while the relation between two displacement vector fields is $\mathbf{\bm{\delta}}_{32}=-\bm{\delta}_{12}$
for the ABC-CBA type LSDW. We further assume that for simplicity the
interlayer potential difference is the same along the layers.

Figure~\ref{fig:6} shows electronic band structures of 3GNRs and
4GNRs with LSDW and their probability distributions for ZLMs. Similarly
to 2GNRs with LSDW, the ZLMs appear not only in ABC-ACB or ABC-CBA
type translation but also in the deviated translations as long as
the translations in the two domains remain in the opposite sides separated
by AAA-AA$'$A line in layer translation vector space. (For the band
structure of AAA and AA$'$A, see Appendix~\ref{app:energy_dispersion}.)
Both zigzag and armchair arrangements show results similar to those
for 2GNR-LSDW. Note that the purple lines in Figs. \ref{fig:6}(c)
and \ref{fig:6}(e) are the interface bound modes \citep{Bi2015,Zarenia2011,Qiao2011}.
These interface bound modes also have probability distributions localized
around the domain wall. However, their energy-momentum dispersion
depends more strongly on the geometry of the interface and electric
field profile than the energy-momentum dispersion of ZLMs does \citep{Qiao2011}.
Moreover, they fade away into the bulk energy levels when a sufficiently
large electric field is applied while the ZLMs survive robustly maintaining
its gapless dispersion.

\begin{figure}[h]
\includegraphics[width=1\linewidth]{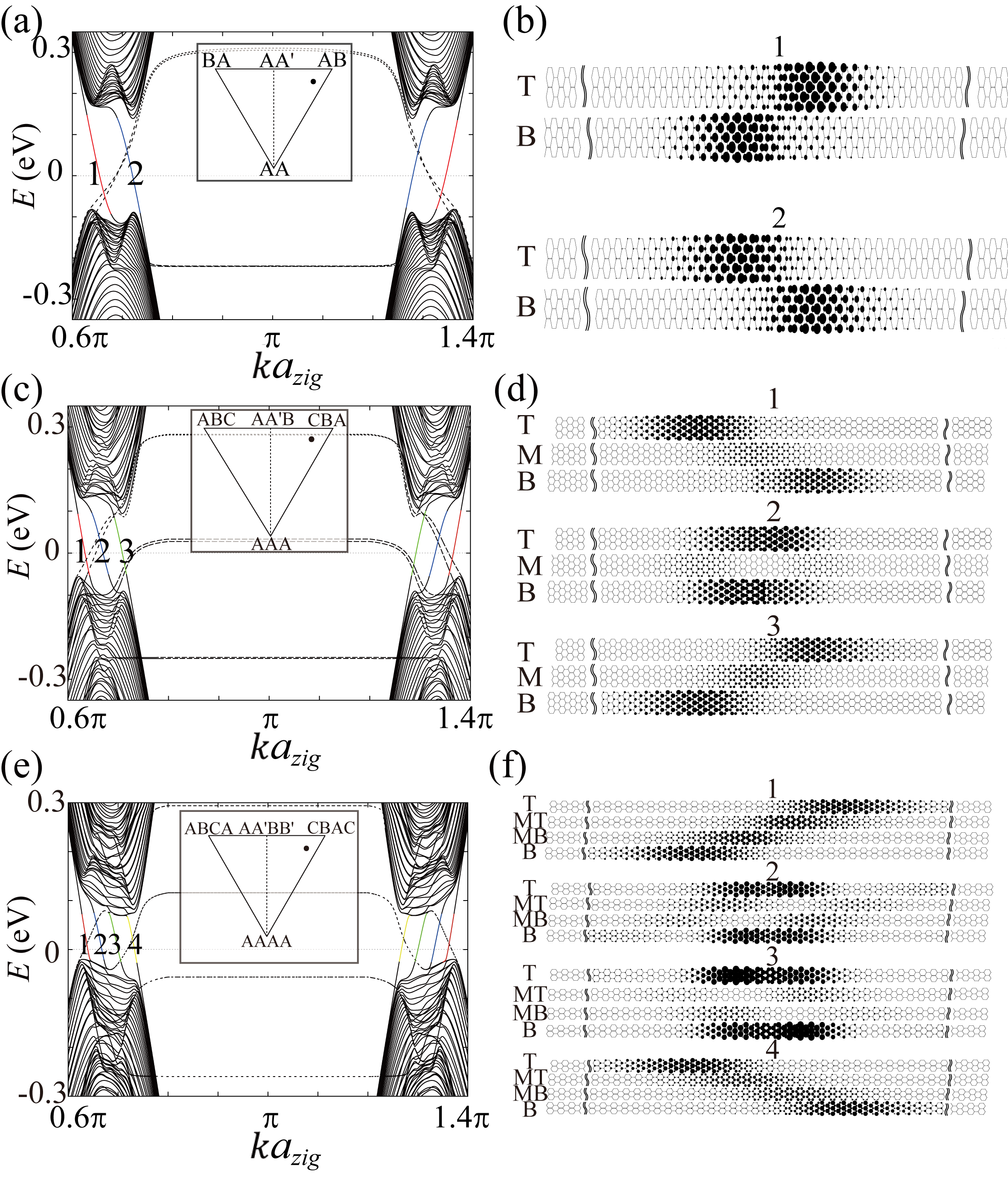} \caption{Band structures of various MGNR-EFDW and corresponding probability
distributions for ZLMs. (a) and (b) 2GNR-EFDW, (c) and (d) 3GNR-EFDW,
and (e) and (f) 4GNR-EFDW, respectively. Owing to time-reversal symmetry,
we only present modes in one valley and omit modes in other valley.
Here, $a_{{\rm zig}}=a$ is the lattice constants of zigzag and armchair
nanoribbons, respectively, and we use the interlayer potential difference
$U=0.5$ eV between the top and bottom layers. }
\label{fig:7} 
\end{figure}

The conclusions that we can draw from our calculations of the LSDW
configurations apply equally for ZLMs that occur in the domain wall
between domains where anti-parallel electric fields are applied. Reversing
the direction of perpendicular electric fields changes the sign of
the valley Chern number, which can be seen in the effective model
in Eq.~(\ref{eq:chiral_gas}). Therefore, like MGNRs with LSDW, there
would be metallic states at the interface between two domains which
have the same stacking structure but opposite field directions. We
call this type of domain wall electric field domain wall (EFDW). Similarly
as LSDW, we construct the EFDW by changing the potential configuration
linearly across the domain wall as 
\begin{equation}
U(x)=\begin{cases}
\frac{U}{2} & (x<0,\mbox{ domain 1}),\\
\frac{U}{2}-\frac{U}{W}x & (0\le x\le W,\mbox{ domain wall}),\\
-\frac{U}{2} & (x>W,\mbox{ domain 2}),
\end{cases}\label{eq:5}
\end{equation}
with the domain wall size $W$. Here we choose the sharp domain wall
with $W=a_{0}$ with abrupt change of the field direction to minimize
the number of unwanted interface bound modes~\cite{Zarenia2011}.
The effect of the domain wall size on ZLMs will be discussed in Sec.
IV.

For Bernal-stacked 2GNRs with EFDW, the properties of the metallic
states have been studied previously in several papers \citep{Zhang2013,Jung2011,Martin2008}.
Here we consider a more generalized setup consisting of stacking faulted
domains due to in-plane layer translation rather than a perfect rhombohedral
stacking, as shown in Fig.~\ref{fig:7} for various MGNRs. In agreement
with prior calculations, we can observe that the number of metallic
states in each valley is always equal to the number of layer $N$,
as expected from Eq.~(\ref{eq:chiral_gas}) for the effective model
which describes a chiral two-dimensional electron gas with chirality
index $N$.

\section{Band gap opening in stacking faulted domains and width of the zero-line
Modes}

\begin{figure}[h]
\includegraphics[width=1\linewidth,height=1.3in]{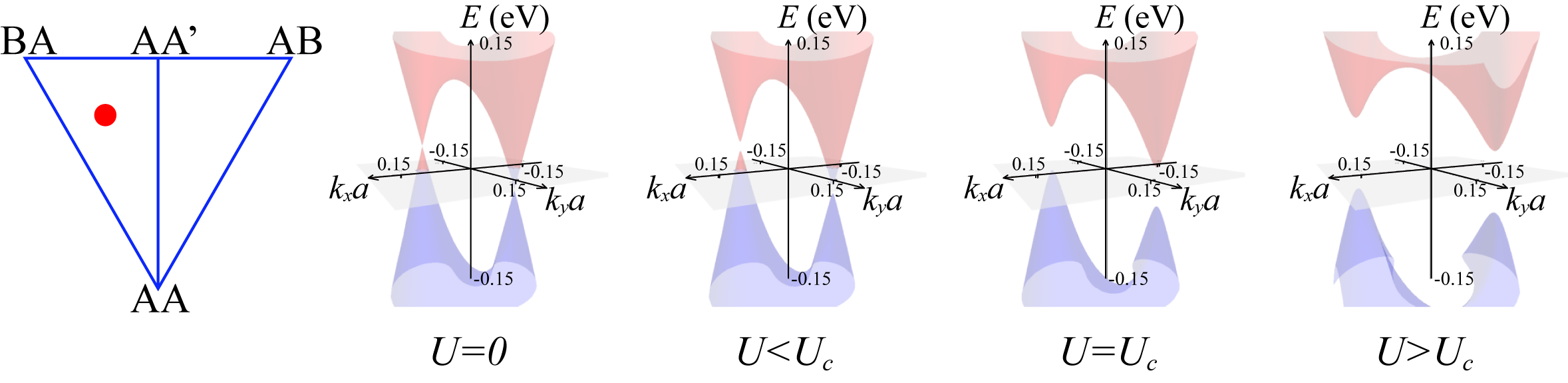} \caption{Illustration for gap opening of bilayer graphene in layer-translated
stacking. Figures from the second panel are the band structures around
the \emph{K} valley with increasing interlayer potential difference
for the stacking with $(\delta_{x},\delta_{y})=(-0.12,0.30)a$ as
represented in the first panel. Electronic states above (below) the
Fermi energy are colored in red (blue).}
\label{fig:8} 
\end{figure}

For practical application of ZLMs in valleytronics devices, it is
required to estimate the proper electric field strength and size of
domain wall for the observation of ZLMs. In this section, we present
numerical results for the electric field strength required to open
an energy gap in stacking faulted domains and the domain wall size
dependence of the ZLM widths. The band gap opening that accompanies
the application of a perpendicular electric field in rhombohedral
graphene is a necessary condition to create the gapped domains flanking
the ZLMs. It was shown that the ability of a perpendicular electric
field to open up a band gap in bilayer graphene will persist for small
departures from the ideal Bernal stacking \citep{Park2015}. This
is true provided that the applied electric field is large enough to
overcome the band asymmetries introduced around the \emph{K} points
due to the stacking fault. Here we present a more detailed account
on the relationship between the required critical perpendicular electric
field for the onset of a band gap in the presence of a stacking fault.
Figure \ref{fig:8} shows a typical example of the onset of band gap
opening in layer translated multilayer graphene where each band structure
represents gradually increasing interlayer potential differences.
When the effective potential between the layers is zero, this layer-translated
bilayer graphene has a metallic band structure with valley degenerate
electron and hole pockets in the Brillouin zone. When the interlayer
potential difference is increased, the bands containing electron (hole)
pockets are raised (lowered) in energy accordingly. When the effective
potential reaches a critical value $U_{{\rm c}}$, the electron and
hole pockets disappear completely and the bilayer graphene becomes
an insulator from this point onwards.

\begin{figure}[h]
\includegraphics[width=1\linewidth]{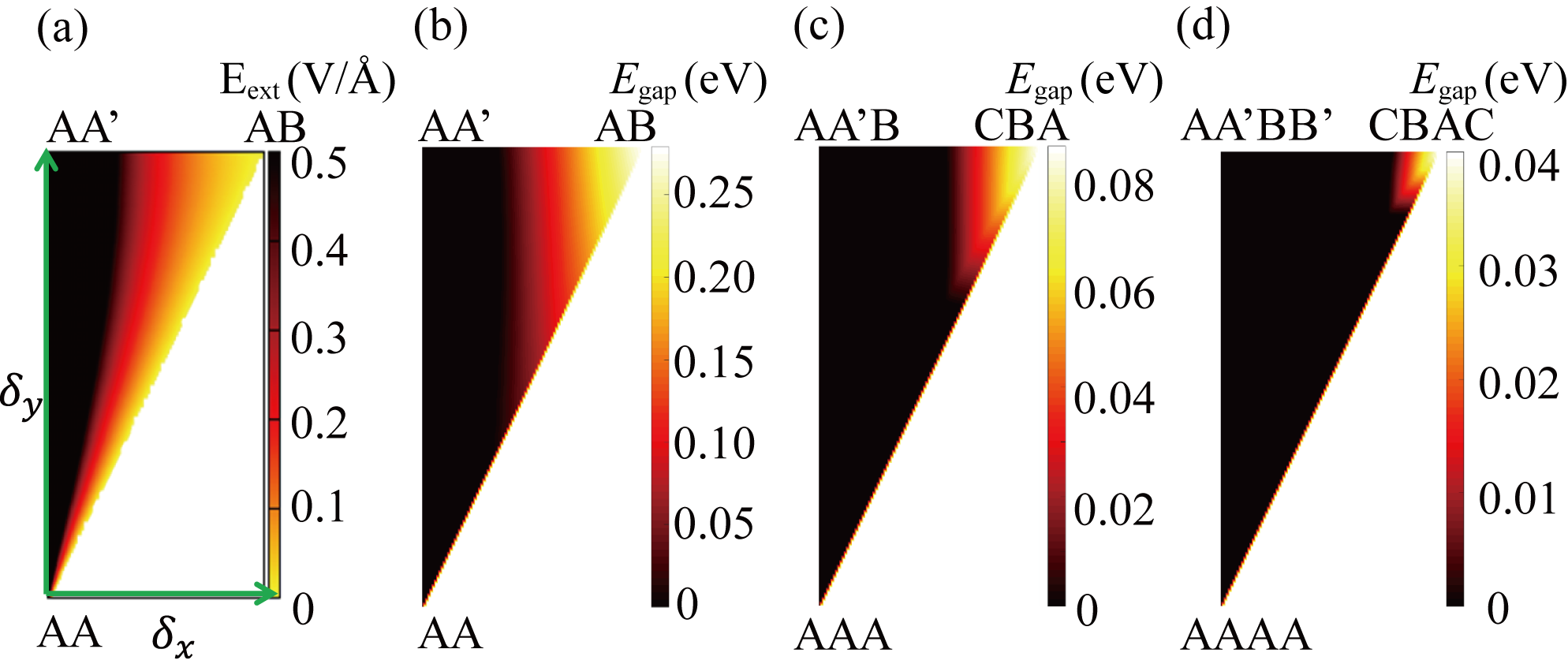} \caption{(a) Minimum external electric field required to make a layer-translated
bilayer graphene insulator is mapped on the irreducible zone of the
layer translation vector space. (b), (c) and (d) are the energy gap
map for bilayer, trilayer and tetralayer graphene with layer-translated
stacking under the external electric field strength $0.5\,\mbox{V}/\textrm{\AA}$.
All the results were obtained using a tight-binding model self-consistent
Hartree calculation.}
\label{fig:9} 
\end{figure}

It is experimentally important to know the critical electric field
required for opening a gap in order to trigger an insulating phase.
Figure \ref{fig:9}(a) shows the critical external electric field
strength for layer-translated bulk bilayer graphene obtained from
a self-consistent Hartree method \cite{Min2007} using a tight-binding
model of Eq.~(\ref{eq:tight_binding}). We obtained an analytic expression
in Eq.~(\ref{eq:A5}) for the critical field required for opening
a band gap in bilayer graphene by solving the roots of the fourth
order polynomial equation of the four-band effective Hamiltonian.
Along the AA-AA$'$ stacking line the critical field becomes infinite
indicating that any stacking along the AA-AA$'$ line remains always
metallic. Note that the topologically distinct domains characterized
by the valley Chern numbers are separated by this AA and AA$'$ line,
as seen in Fig.~\ref{fig:4}.

The magnitude of the achievable gap depends on the particular stacking
configuration. In Figs.~\ref{fig:9}(b)$\sim$(d) we show the band
gaps that develop under an external electric field of $0.5\,\mathrm{V/\mathring{A}}$
as a function of layer translation, in which we clearly see a decrease
in the gap size as we move closer to the AA-AA$'$ line for bilayer
graphene (AAA-AA$'$A line for trilayer graphene, AAAA-AA$'$AA$'$
line for tetralayer graphene). As discussed earlier, this gap decrease
has its origin in the band structure of the layer-translated multilayer
graphene. By means of a self-consistent Hartree screening calculation,
we verified that the effective potential difference between layers
under electric fields is almost independent of stacking configuration
in bilayer graphene. For example, the potential difference $U$ between
layers in bilayer graphene under electric fields of 0.5 $\mbox{V}/\textrm{\AA}$
turns out to be about 0.63 eV for all layer translated stackings.
This result partially justifies the validity of our assumption of
constant effective interlayer potentials in our calculation for MGNR-LSDW,
though the stacking dependence of the interlayer potentials becomes
more important as the number of layers increases.

\begin{figure}
\includegraphics[width=1\linewidth]{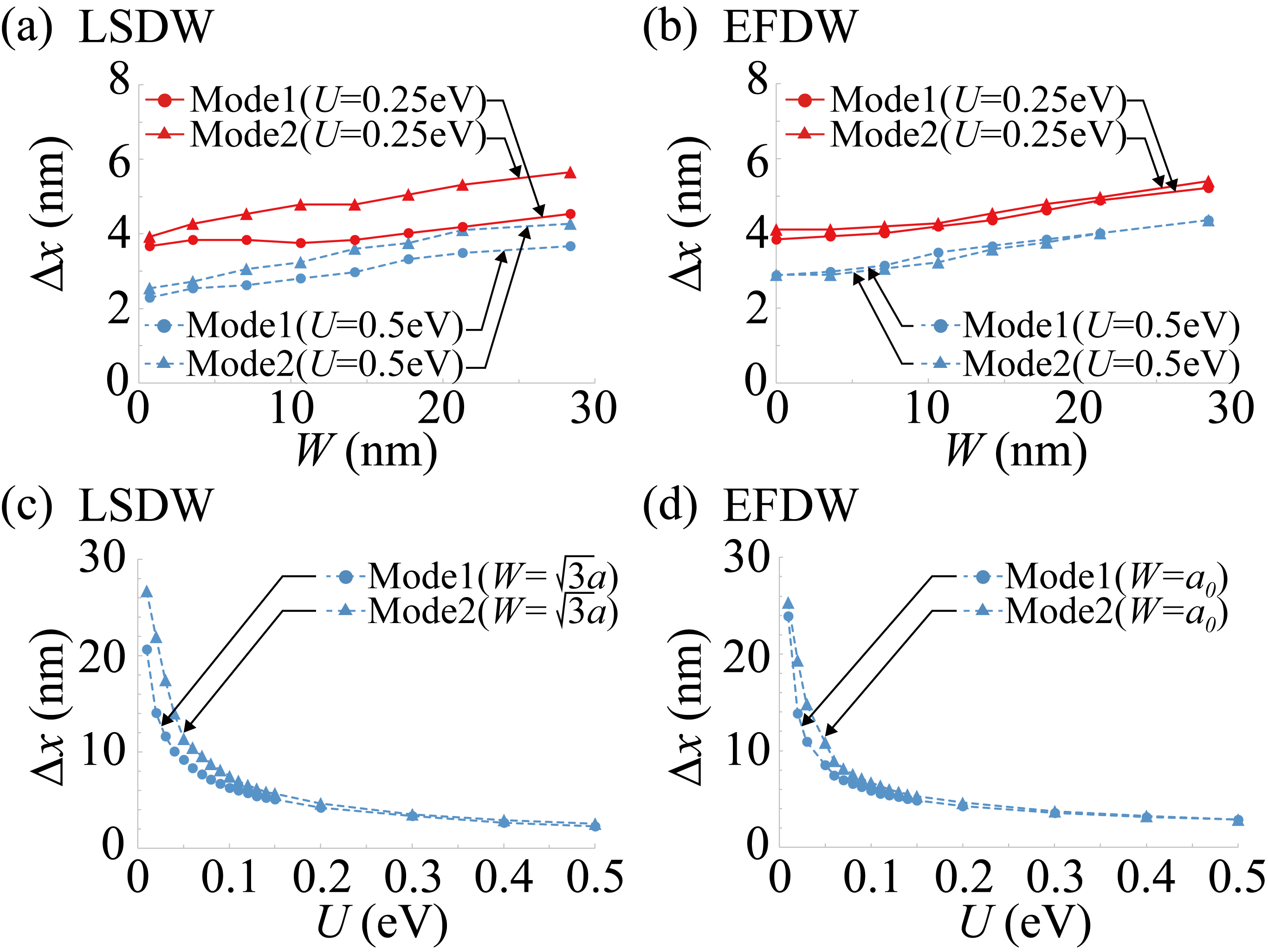} \caption{(a) and (b): Size of ZLM $\Delta x$ as a function of domain wall
width $W$ for (a) LSDW and (b) EFDW for $U=0.25$ eV (red solid lines)
and $U=0.5$ eV (blue dashed lines). Triangles and circles represent
two ZLMs 1 and 2, respectively, in Figs.~\ref{fig:4} and \ref{fig:7}.
(c) and (d): Size of ZLM $\Delta x$ as a function of potential strength
$U$ for (c) LSDW and (d) EFDW for sharp domain wall with $W=\sqrt{3}a$
for LSDW and $W=a_{0}$ for EFDW, respectively.}
\label{fig:10} 
\end{figure}

In order to provide a practical guidance for experiments aiming to
probe ZLM local density of states, we analyzed the influence of the
domain wall size $W$ and the magnitude of the potential difference
$U$ on the width of ZLM $\Delta x$ defined as the distance that
contains 90\% of probability distribution centered inside the domain
wall. In Figs.~\ref{fig:10}(a) and \ref{fig:10}(b), the ZLM width
$\Delta x$ versus the domain wall width $W$ is illustrated for LSDW
and EFDW, respectively, for the potential difference $U=0.25$ eV
and $U=0.5$ eV. We find that even in the sharp domain wall ($W=\sqrt{3}a$)
the ZLMs have a finite size, and the size increases relatively weakly
as the domain wall width increases. Figures \ref{fig:10}(c) and \ref{fig:10}(d)
show the potential strength dependence of the ZLM size for LSDW and
EFDW, respectively, with $W=\sqrt{3}a$ and $W=a_{0}$. In general,
the width of ZLMs $\Delta x$ decreases rapidly as the potential difference
$U$ increases for both LSDW and EFDW.

\section{Summary and conclusion}

In this work we have studied the zero-line modes (ZLMs) in multilayer
graphene nanoribbon geometries in the presence of domains whose interlayer
stacking deviates from perfect chiral stacking, namely the Bernal
alignment for bilayer and rhombohedral stacking for multilayers, providing
a systematic and comprehensive picture on the existence of ZLMs in
arbitrarily displaced multilayer graphene. The analysis carried out
in this work expands the validity of the analysis of the ZLMs based
on the valley Hall conductivity of the insulating domains for chirally
stacked multilayer graphene and confirms the possibility of realizing
the ZLMs in actual experimental devices where stacking faults may
be present either due to unwanted disorder effects or artificial strain
fields. We have discussed the conditions of stacking order in the
bulk and external perpendicular electric fields required for opening
the band gap in the bulk necessary to generate the ZLMs in the system.
We can conclude that stacking faulted domains that deviate from perfect
rhombohedral stacking can also host ZLMs provided that sufficiently
large perpendicular electric fields can be applied to open up a band
gap in the bulk. The valley Chern number signs for the bulk are defined
by stacking configurations that can be classified as AB-like and BA-like
regions and are divided by the AA-AA$'$ stacking lines. We found
that a larger electric field is required to open up a band gap when
the perfect chiral stacking is modified through in-plane sliding.
This critical field has been obtained both analytically and numerically
for a variety of stacking configurations. The explicit calculations
of the ZLM probability distribution for both layer stacking and electric
field domain walls with zigzag and armchair edge alignments provide
information that can be useful when designing junctions made of intersecting
ZLMs whose current partitioning properties are expected to depend
on the overlap between the incoming and outgoing modes \citep{Qiao2014,Ezawa2013}.
Our work also provides insight for understanding the ZLMs that can
be expected in layer misaligned systems such as twisted bilayer graphene
and multilayers with tilted boundaries where spatially varying local
stacking configurations are present. 
\begin{acknowledgments}
This research was supported by Basic Science Research Program through
the National Research Foundation of Korea (NRF) funded by the Ministry
of Education (MOE) of Korea under Grant No. 2015R1D1A1A01058071. G.K.
thanks the financial support from the Priority Research Center Program
(2010-0020207), and the Basic Science Research Program (2013R1A2009131)
through NRF funded by the MOE. J.J. acknowledges financial support
from the Korean NRF through the grant NRF-2016R1A2B4010105. 
\end{acknowledgments}

\appendix

\section{Critical potential difference in layer-translated bilayer graphene\label{app:Critical-potential-difference}}

The Hamiltonian for bilayer graphene with arbitrary layer translation
in a continuum model can be written as\citep{Koshino2013} 
\begin{equation}
H=\left(\begin{array}{cccc}
\frac{U}{2} & \pi^{\ast} & u\left(\bm{\delta}\right) & w\left(\bm{\delta}\right)\\
\pi & \frac{U}{2} & v\left(\bm{\delta}\right) & u\left(\bm{\delta}\right)\\
u^{\ast}\left(\bm{\delta}\right) & v^{\ast}\left(\bm{\delta}\right) & -\frac{U}{2} & \pi^{\ast}\\
w^{\ast}\left(\bm{\delta}\right) & u^{\ast}\left(\bm{\delta}\right) & \pi & -\frac{U}{2}
\end{array}\right),\label{eq:A1}
\end{equation}
where $\pi=\hbar v(k_{x}+ik_{y})$ and interlayer hopping terms are
given by 
\begin{align}
u\left(\bm{\delta}\right)= & u_{_{A_{1}A_{2}}}\left(\bm{\delta}\right)=u_{_{B_{1}B_{2}}}\left(\bm{\delta}\right)\nonumber \\
= & \frac{\gamma}{3}\left[1+2\cos\left(\frac{2\pi\delta_{x}}{3a_{0}}\right)e^{\frac{2\pi i\xi\delta_{y}}{\sqrt{3}a_{0}}}\right],\\
v\left(\bm{\delta}\right)= & u_{_{A_{2}B_{1}}}\left(\bm{\delta}\right)\nonumber \\
= & \frac{\gamma}{3}\left[1+2\cos\left(\frac{2\pi}{3}\left(\frac{\delta_{x}}{a_{0}}+1\right)\right)e^{\frac{2\pi i\xi\delta_{y}}{\sqrt{3}a_{0}}}\right],\\
w\left(\bm{\delta}\right)= & u_{_{B_{2}A_{1}}}\left(\bm{\delta}\right)\nonumber \\
= & \frac{\gamma}{3}\left[1+2\cos\left(\frac{2\pi}{3}\left(\frac{\delta_{x}}{a_{0}}-1\right)\right)e^{\frac{2\pi i\xi\delta_{y}}{\sqrt{3}a_{0}}}\right],
\end{align}
where $\xi=\pm1$ for K, K$'$ valleys and $\bm{\delta}=(\delta_{x},\delta_{y})$
is the layer translation vector defined in Fig.~\ref{fig:1} with
$\bm{\delta}=0$ for AA stacking. The above expression for interlayer
hopping terms is obtained through Fourier transformation at the corner
of the first Brillouin zone\cite{Koshino2013} and resembles the stacking-dependent
interlayer interaction in bilayer graphene given in Ref.~\citealp{Jung2014moire}.
Then the required effective potential to open energy gap can be obtained
as 
\begin{equation}
\frac{U_{{\rm c}}}{2}=\frac{\sqrt{\left(\left|u\right|^{2}+\left|vw\right|^{2}\right)^{2}-\left|u^{2}-vw\right|^{2}}}{\left|\left|v\right|-\left|w\right|\right|}.\label{eq:A5}
\end{equation}
Note that stackings along AA-AA$'$ line corresponding to $|v|=|w|$
require an infinite potential difference for gap opening. This means
that bilayer graphene in theses stackings can not be insulators by
applying a perpendicular electric field within the continuum model
approximation. However, when two Dirac cones at the K and K' valleys
are directed toward the M point and meet each other, it is possible
that stacking along AA-AA' line can be gapped even at a large but
finite potential energy difference.

\section{Energy dispersions in various layer-translated multilayer graphene}

\label{app:energy_dispersion} In this appendix, we illustrate various
atomic structures and corresponding energy dispersions for the layer-translated
multilayer graphene configurations labeled as AA, AA$'$, AAA, AAAA,
AA$'$A and AA$'$AA$'$ and discussed in the main text.

\begin{figure*}[t]
\includegraphics[width=1\linewidth]{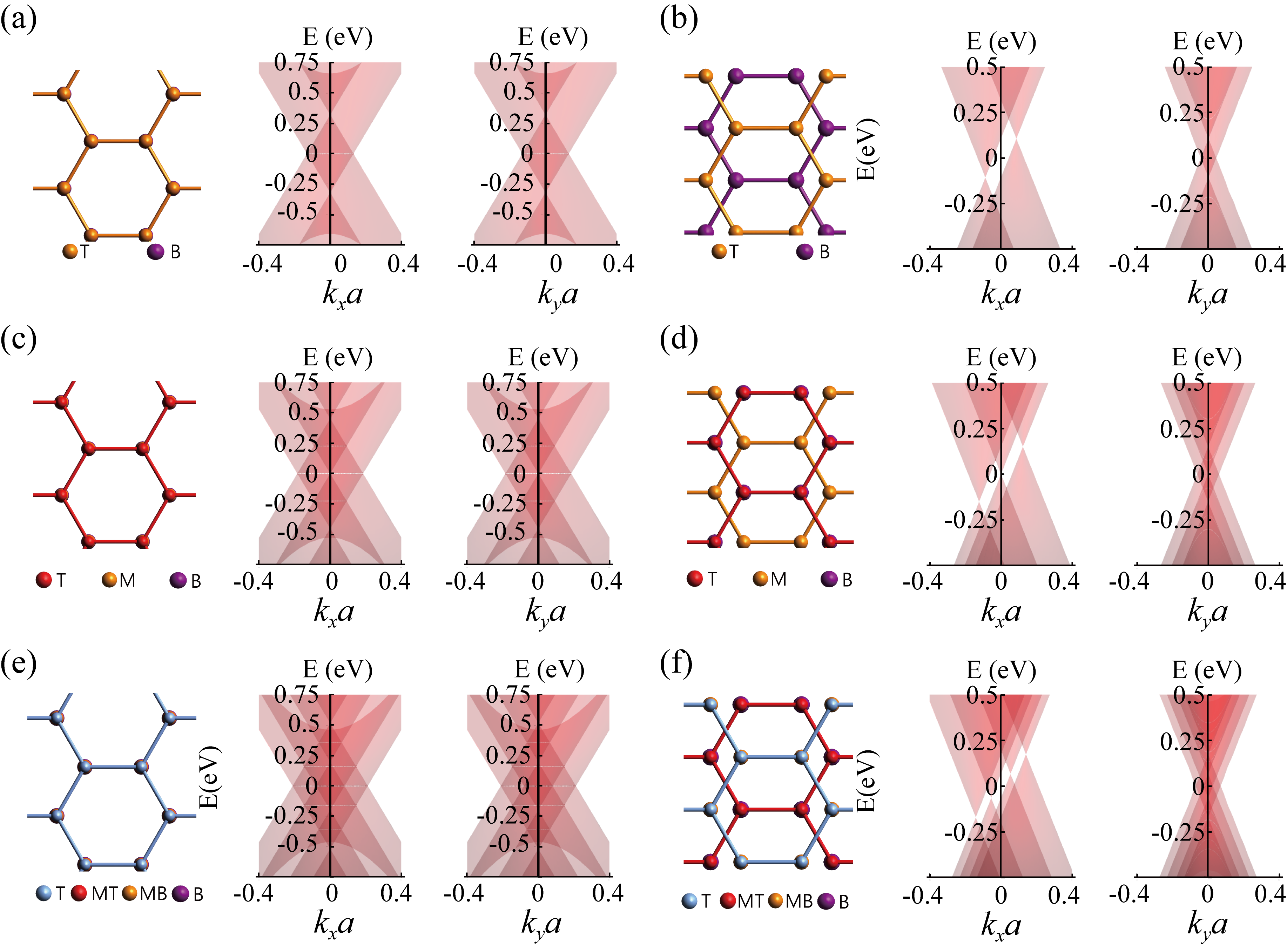} \caption{Several atomic structures and corresponding energy dispersions for
(a) AA, (b) AA$'$, (c) AAA, (d) AA$'$A, (e) AAAA and (f) AA$'$AA$'$
stackings.}
\end{figure*}

  \bibliographystyle{apsrev4-1}
\nocite{*}
\bibliography{myref}

\end{document}